\documentclass[twocolumn,aps,pra,10pt,showpacs]{revtex4-1}
\usepackage{bm}
\usepackage{amsfonts}
\usepackage{amssymb}
\usepackage{amsmath}
\usepackage{graphicx}
\newcommand{\sint}{\;\int\mspace{-26mu}\sum}
\begin{document}

\title{Three measures of fidelity for photon states}
\author{Iwo Bialynicki-Birula}\email{birula@cft.edu.pl}
\affiliation{Center for Theoretical Physics, Polish Academy of Sciences\\
Aleja Lotnik\'ow 32/46, 02-668 Warsaw, Poland}
\author{Zofia Bialynicka-Birula}
\affiliation{Institute of Physics, Polish Academy of Sciences\\
Aleja Lotnik\'ow 32/46, 02-668 Warsaw, Poland}

\begin{abstract}
We show that the standard method of introducing the quantum description of the electromagnetic field -- by canonical field quantization -- is not the only one. We have chosen here the relativistic quantum mechanics of the photon as the starting point. The treatment of photons as elementary particles merges smoothly with the description in terms of the quantized electromagnetic field but it also reveals some essential differences. The most striking result is the appearance of various measures of fidelity for quantum states of photons. These measures are used to characterize the localization of photons.
\end{abstract}

\maketitle

\section{Introduction}

In his fundamental paper \cite{pamd} on the quantum theory of electromagnetic radiation Dirac wrote
``There is thus a complete harmony between the wave and light-quantum descriptions\dots'' and he continued,
``We shall actually build up the theory from the light-quantum point of view\dots''. Our present understanding of the nature of photons significantly differs from what has been known years ago when the concept of a photon has only been emerging. These changing views were described in a vivid way by Shore \cite{bws}. The theoretical tools that we have presently at our disposal enable us to formulate the quantum theory of photons more completely than Dirac could do. Many papers were written on this subject so that it came as a surprise to us that an important element of the quantum theory of photons was overlooked: the {\em ambiguity} in the probabilistic interpretation of photons states embodied in the modern notion of fidelity.

Fidelity of photon states was the subject of many experimental and theoretical studies. However, in these studies only the degree of freedom connected with polarization was taken into account. In this work we use the full description of photon states based on quantum mechanics of photons described in our previous publications  \cite{app,pwf,hur,rs,qm}. We shall show that there are two equally sensible definitions of fidelity, giving different results for the same pair of states.

\section{Quantum mechanics of photons --- Momentum representation}\label{momr}

Considering the fact that a photon is an elementary particle, one may expect that it is possible to formulate the quantum theory of photon patterned after the quantum mechanics of massive particles. Vanishing of the photon rest mass creates, however, serious difficulties in the construction of such a theory. This is the reason why quantum mechanics of photon is not widely known. We shall use here a complete version of this theory.

On general grounds one should expect that the state of the photon is described by a wave function, as is the case for all other quantum particles. In contrast to massive particles, the photon wave function in {\em momentum representation}, and not in position representation, plays the fundamental role. Since there are two types of photons, with positive helicity (right-handed) and with negative helicity (left-handed), we need two wave functions $f_+(\bm k)$ and $f_-(\bm k)$, where ${\bm k}$ is the wave vector. This basic property of photons is completely lost in the standard formulation where all possible polarizations, (circular, vertical, horizontal, etc.) are treated on equal footing.

Distinctness of photons with positive and negative helicity shows up in the transformation properties of the wave functions $f_\pm(\bm k)$. According to the Wigner classification \cite{wig} these functions form two {\em different} one-dimensional representations of the Poincar\'e group. This is why one cannot directly superpose the wave functions $f_+(\bm k)$ and $f_-(\bm k)$.

The wave functions $f_\pm(\bm k)$ have the standard probabilistic interpretation: the modulus squared of the wave function determines the probability density to find the photon with the momentum $\hbar{\bm k}$.

A general (pure) photon state is described by the following two-component wave function:
\begin{align}\label{pwf}
{\mathfrak{\bm{f}}({\bm k})}=\left(\begin{array}{c}f_+({\bm k})\\f_-({\bm k})\end{array}\right).
\end{align}
We shall use, sometimes, a different notation introducing a function $f({\bm k},\lambda)$ where the parameter $\lambda$ takes on two values $\pm$.

The photon wave functions (\ref{pwf}) form Hilbert space endowed with the inner product:
\begin{align}\label{inpr}
\langle{\mathfrak{\bm{f}}_1}|{\mathfrak{\bm{f}}_2}\rangle
=\int\!\frac{d^3k}{k}{\mathfrak{\bm{f}}_1^*({\bm k})}{\mathfrak{\bm{f}}_2({\bm k})}=\sum_\lambda\frac{d^3k}{k}f^*_1({\bm k},\lambda)f_2({\bm k},\lambda),
\end{align}
where $\displaystyle\sint$ denotes the integration over $\bm k$ and the summation over $\lambda$. The norm induced by this inner product leads to the following normalization condition:
\begin{align}\label{norm}
\int\!\frac{d^3k}{k}|{\mathfrak{\bm{f}}({\bm k})}|^2=1.
\end{align}
The appearance of the length of the wave vector in the denominator is required by relativistic invariance.

Photons are relativistic particles. Therefore, in their description the essential role is played by the generators of the Poincar\'e group \cite{qed}. These generators are Hermitian with respect to the inner product (\ref{inpr}). Therefore, the Poincar\'e transformations do not change the norm and the transformed wave functions can only pick up a phase factor,
\begin{align}\label{phf}
f'_\pm({\bm k}')=e^{\pm i\Theta({\bm k})}f_\pm({\bm k}).
\end{align}
The sign difference in this formula distinguishes photons with opposite helicities. We shall show in the Appendix how to determine the phase $\Theta({\bm k})$ for a given Poincar\'e transformation. The transformation properties of photon wave functions show that one should not treat the two helicity states as analogous to two spin components of the electron wave function. Under rotations (and also under Lorentz transformations) the relative proportions of the spin up and spin down do change, while the helicity components only change their phase.

\section{Quantum mechanics of photons --- Position representation}\label{posr}

The concept of the photon wave function in position representation has a long history described in detail in \cite{pwf}. It has been recognized from the very beginning that this function cannot have all the expected properties and for that reason this concept has not been widely accepted. One of the reasons is that the multiplication by $\bm r$ cannot be used as the position operator. Another reason is the absence of the local probability density and the corresponding probability current.

In nonrelativistic quantum mechanics there are no difficulties with the transformation of the wave functions from  momentum to position representation. This is accomplished with the use of the straightforward Fourier transformation which defines the connection between the two representations. This simple prescription applied to photon wave function in momentum representation also gives some kind of a position wave function satisfying the d'Alembert wave equation,
\begin{align}\label{wf}
\phi_\pm({\bm r},t)=\int\frac{d^3k}{(2\pi)^{3/2}k}f_\pm(\bm k)e^{-i\omega t+i{\bm k}\cdot{\bm r}}.
\end{align}
However, as has been discovered by Pauli \cite{pauli}, the wave functions $\phi$ are unacceptable because they are nonlocal. Namely, after a rotation or a Lorentz boost applied to $f_\pm(\bm k)$ the transformed wave function $\phi'$ depends on the values of $\phi$ {\em at all points}.

A resolution of this dilemma was based in \cite{pwf} on the fact that photons are, after all, associated with the electromagnetic field and the electromagnetic field has well defined local transformation properties. The task, therefore, was to find a Fourier-type relation between the momentum wave functions and the electromagnetic field. In order to complete this task we used the construction of Whittaker \cite{whitt}, who discovered that one may construct the electromagnetic field satisfying Maxwell equations from the derivatives of a solution (\ref{wf}) of the d'Alembert equation. The complexified version of the Whittaker construction gives the complex combination of the electric and magnetic field (named the Riemann-Silberstein (RS) vector in \cite{pwf,rs}) in the form:
\begin{align}\label{whitt}
{\bm F}(\bm r,t)\!=\!\frac{{\bm D}(\bm r,t)}{\sqrt{2\epsilon_0}}+i\frac{{\bm B}(\bm r,t)}{\sqrt{2\mu_0}}\!=\!\left[\!\begin{array}{c}
\partial_x\partial_z+i/c\,\partial_y\partial_t\\
\partial_y\partial_z-i/c\,\partial_x\partial_t\\
-\partial_x^2-\partial_y^2
\end{array}
\!\right]\!\chi({\bm r},t),
\end{align}
where $\chi({\bm r},t)$ is any solution of the d'Alembert equation. The RS vector obeys the Maxwell equations in the form of the Schr\"odinger-type equation,
\begin{align}\label{sch}
i\partial_t{\bm F}({\bm r},t)=c{\bm\nabla}\times{\bm F}({\bm r},t),\quad {\bm\nabla}\!\cdot\!{\bm F}({\bm r},t)=0.
\end{align}

In order to connect the photon wave functions with the electromagnetic field we start with the general solution of the Maxwell equations for the RS vector written as the superposition of plane waves,
\begin{align}\label{genpwf}
&{\bm F}({\bm r},t)=\sqrt{\hbar c}\int\!\!\frac{d^3k}{(2\pi)^{3/2}}{\bm e}(\bm k)\nonumber\\
&\times\left[f_+(\bm k)e^{-i\omega t+i{\bm k}\cdot{\bm r}}+f^*_-(\bm k)e^{i\omega t-i{\bm k}\cdot{\bm r}}\right].
\end{align}
where $f_+(\bm k)$ and $f^*_-(\bm k)$ are two complex functions and
\begin{align}\label{ee}
{\bm e}(\bm k)=\frac{1}{\sqrt{2k^2(k_x^2+k_y^2)}}
\left[\begin{array}{c}
-k_xk_z+i k_y k\\
-k_yk_z-i k_x k\\
k_x^2+k_y^2
\end{array}\right].
\end{align}
The coefficient $\sqrt{\hbar c}$ is chosen so that after second quantization the electromagnetic field operator is properly normalized.

We identify the functions $f_+(\bm k)$ and $f^*_-(\bm k)$ with the photon wave functions in momentum representation. However, since the wave functions of particles may contain only positive frequencies, we identify $f_+(\bm k)$ with the positive helicity photon wave function and $f^*_-(\bm k)$ with the complex conjugate of the negative helicity photon wave function. This identification is consistent with the transformation properties of the electromagnetic field, as is explained in the Appendix.

The two parts of ${\bm F}(\bm r,t)={\bm\Psi}_+({\bm r},t)+{\bm\Psi}_-^*({\bm r},t)$ are natural candidates for the photon wave functions with positive and negative helicity in position representation,
\begin{align}\label{pwf1}
{\bm\Psi}_\pm({\bm r},t)=\sqrt{\hbar c}\int\frac{d^3k}{(2\pi)^{3/2}}{\bm e}_\pm(\bm k)f_\pm(\bm k)e^{-i\omega t+i{\bm k}\cdot{\bm r}},
\end{align}
where ${\bm e}_+(\bm k)={\bm e}(\bm k)$ and ${\bm e}_-(\bm k)={\bm e}^*(\bm k)$. It may seem that photons with positive helicity are distinguished since the wave function with negative helicity appears as the complex conjugate. This is, however, the result of an arbitrary choice (for historical reasons) of the sign of the imaginary part in the definition (\ref{whitt}) of the RS vector. The remaining technical problem is that of normalization. Photon wave functions must be properly normalized while the classical electromagnetic field can have arbitrarily large amplitude. This issue will be resolved in Sec.~\ref{qft} with the introduction of quantum field operators.

The physical meaning of the position wave function follows from Glauber's theory of optical coherence \cite{roy0}. Namely, the modulus squared of this function determines ``the probability per unit time that a photon be absorbed by an ideal detector at point $\bm r$ at time $t$''.

\section{Photons --- Second quantization}\label{phsq}

In most situations we are dealing with the states that contain huge numbers of photons. Therefore, we need a formalism which would not concentrate on single photons but would allow for an efficient description of many photons. This formalism is based on the {\em second quantization}. By invoking directly second quantization, we bypass all the steps made in the standard approach (introduction of the potentials, gauge- fixing, expansion of the electromagnetic field into monochromatic modes, etc.).

According to the rules of second quantization  \cite{dqm,ls} the photon wave functions are replaced by the annihilation operators $a({\bm k},\lambda)$, while complex conjugate wave functions are replaced by the creation operators $a^\dagger({\bm k},\lambda)$,
\begin{align}\label{sq}
f({\bm k},\lambda)\to a({\bm k},\lambda),\quad f^*({\bm k},\lambda)\to a^\dagger({\bm k},\lambda).
\end{align}
\begin{align}\label{regcom}
[a({\bm k},\lambda),a^\dagger({\bm k}',\lambda')]=\delta_{\lambda\lambda'}k\delta^{(3)}({\bm k}-{\bm k}').
\end{align}
The factor of $k$ on the right hand side follows from the relativistic normalization (\ref{norm}) of the photon wave functions. The creation operators  $a^\dagger_f$ of physically realizable (normalizable wave packets) photon states are superpositions of the operators $a^\dagger({\bm k}',\lambda)$,
\begin{align}\label{cr}
a^\dagger_f=\sint\frac{d^3k}{k} f({\bm k},\lambda)a^\dagger({\bm k},\lambda).
\end{align}

The complete space of states (the Fock space) contains  all superpositions of $N$-photon states. The distinguished role among them is played by {\em coherent states} \cite{roy}. They describe, in a good approximation, light emitted by lasers, klystrons or radio stations. Coherent states are superpositions of the states with different numbers of {\em identical} photons,
\begin{align}\label{coh}
|{\rm coh}\rangle=
=e^{-\langle N\rangle/2}\exp\left(\langle N\rangle^{1/2}a^\dagger_f\right)|0\rangle.
\end{align}
Each photon is in the state described by the wave function $f({\bf k},\lambda)$ and $\langle N\rangle$ is the average number of photons in the coherent state.

\section{Operator of the quantized electromagnetic field}\label{qft}

The field operator ${\hat{\bm F}}(\bm r,t)$ of the quantized electromagnetic field obtained according to the rules of second quantization is:
\begin{align}\label{fin1}
&{\hat{\bm F}}(\bm r,t)=\sqrt{\hbar c}\int\!\frac{d^3k}{(2\pi)^{3/2}}{\bm e}(\bm k)\nonumber\\
&\times\left[a({\bm k},+)e^{i\bm k\cdot\bm r-i\omega t}+a^\dagger({\bm k},-)e^{-i\bm k\cdot\bm r+i\omega t}\right].
\end{align}
The direct connections of the electromagnetic field operator with the classical electromagnetic field can be found in two ways. First, the average value of the quantum field in a coherent state (\ref{coh}) is proportional to the electromagnetic field (\ref{genpwf}) which is associated with photons populating the coherent state,
\begin{align}\label{avcoh}
\langle{\rm coh}|{\hat{\bm F}}(\bm r,t)|{\rm coh}\rangle=\langle N\rangle^{1/2} {\bm F}({\bm r},t).
\end{align}
The proportionality constant is the square root of the average number of photons in the coherent state. Second, the sum of the matrix elements of the field operator between one-photon states of both helicities and the vacuum state gives the RS vector associated with the photon wave function,
\begin{align}\label{mel}
\langle 0|{\hat{\bm F}}(\bm r,t)|f_+\rangle+\langle f_-|{\hat{\bm F}}(\bm r,t)|0\rangle ={\bm F}({\bm r},t).
\end{align}

\section{Fidelity --- Measure of localization}\label{fidel}

The standard definition of fidelity $\mathcal{F}$ applied to two states of the photon, described by the wave functions in momentum space, gives
\begin{align}\label{fidm}
\mathcal{F}_{\rm m}
=\frac{\left|\langle{\mathfrak{\bm{f}}_1}|
{\mathfrak{\bm{f}}_2}\rangle\right|^2}
{\langle{\mathfrak{\bm{f}}_1}|
{\mathfrak{\bm{f}}_1}\rangle\langle{\mathfrak{\bm{f}}_2}|
{\mathfrak{\bm{f}}_2}\rangle},
\end{align}
where the inner product is defined in (\ref{inpr}) and we allowed for unnormalized wave functions. The subscript ``m'' stands for momentum representation. The same measure of fidelity is obtained for one photon states $|1_f\rangle$ in Fock space created from the vacuum by the operators (\ref{cr}),
\begin{align}\label{fidfock}
\mathcal{F}_{\rm m}
=|\langle 1_{f_1}|1_{f_2}\rangle|^2.
\end{align}

In the language of the standard quantum theory, fidelity is just \cite{gt} ``The probability that a system $S$ in the state $|\psi\rangle$ will be found to be in an arbitrary state $|\phi\rangle$.'' The overlap of the two wave functions $|\langle\phi|\psi\rangle|^2$ is the measure of their similarity. In nonrelativistic quantum mechanics, and also in any abstract quantum theory, this measure of the similarity between two quantum states is unique because the inner product is unique. We will argue here that for photons this uniqueness is questionable. The formula (\ref{fidm}) rewritten in terms of photon wave functions in position representation takes a nonlocal form,
\begin{align}\label{sprf}
\langle{\mathfrak{\bm{f}}_1}|
{\mathfrak{\bm{f}}_2}\rangle&=\sint\!\frac{d^3k}{k}f_{1\lambda}^*({\bm k})f_{2\lambda}({\bm k})\nonumber\\
&=\frac{1}{2\pi^2}\sum_\lambda\iint\!\frac{d^3rd^3r'}{|\bm r-\bm r'|^2}{\bm \Psi}^*_{1\lambda}(\bm r,t)\!\cdot\!{\bm \Psi}_{2\lambda}(\bm r',t).
\end{align}
The norm in momentum space expressed in terms of the photon wave function in position space,
\begin{align}\label{normf}
\langle{\mathfrak{\bm{f}}}|
{\mathfrak{\bm{f}}}\rangle=\frac{1}{2\pi^2}\sum_\lambda\iint\!\frac{d^3rd^3r'}{|\bm r-\bm r'|^2}{\bm \Psi}^*_{\lambda}(\bm r,t)\!\cdot\!{\bm \Psi}_{\lambda}(\bm r',t),
\end{align}
has the physical interpretation of the total number of photons \cite{pwf,zeld}. The normalization condition $\langle{\mathfrak{\bm{f}}}|
{\mathfrak{\bm{f}}}\rangle=1$ can be interpreted in position representation as a requirement that the corresponding electromagnetic field contains on average only one photon.

Having the wave functions in position representation at our disposal, we may introduce the Hilbert space based on the following local form of the inner product:
\begin{align}\label{locinpr}
\langle{\bm \Psi}_1|{\bm \Psi}_2\rangle=\sum_\lambda\int\!d^3r{\bm \Psi}^*_{\lambda 1}(\bm r,t)\!\cdot\!{\bm \Psi}_{\lambda 2}(\bm r,t).
\end{align}
This new inner product enables us to introduce the fidelity in position representation,
\begin{align}\label{fidp}
\mathcal{F}_{\rm p}
=\frac{\left|\langle{\bm \Psi}_1|{\bm \Psi}_2\rangle\right|^2}
{\langle{\bm \Psi}_1|{\bm \Psi}_1\rangle\langle{\bm \Psi}_2|{\bm \Psi}_2\rangle}.
\end{align}
Two measures of fidelity, $\mathcal{F}_{\rm m}$ and $\mathcal{F}_{\rm p}$, give different values for the same photon states, as we show below. In contrast, for relativistic electrons the momentum and position fidelities are equal. This difference underscores the peculiarity of the photon quantum mechanics.

There is another difference between the momentum and position fidelities for photons. Both fidelities are invariant under rotations but under Lorentz transformations $\mathcal{F}_{\rm p}$ changes while $\mathcal{F}_{\rm m}$ is invariant. However, even for transformations with velocities close to the speed of light, these changes are much smaller than the difference between $\mathcal{F}_{\rm m}$ and $\mathcal{F}_{\rm p}$.

The fidelities $\mathcal{F}_{\rm m}$ and $\mathcal{F}_{\rm p}$ are measures of the similarity between quantum photon states. One may, however, introduce yet another measure of fidelity that is related to the properties of the electromagnetic fields associated with photons. This definition employs coherent states. Fidelity of coherent states $\mathcal{F}_{\rm c}$ is defined with the use of the scalar product in Fock space,
\begin{align}\label{fidc}
\mathcal{F}_{\rm c}
=\frac{\left|\langle{\rm coh}_1|{\rm coh}_2\rangle\right|^2}
{\langle{\rm coh}_1|{\rm coh}_1\rangle\langle{\rm coh}_2|{\rm coh}_2\rangle}.
\end{align}
Since coherent states are defined in (\ref{coh}) in terms of the photon wave function in momentum representation, there should exist a relation between $\mathcal{F}_{\rm c}$ and $\mathcal{F}_{\rm m}$. This relation depends not only on the fidelity but also on the phase difference $\varphi$ of two photon wave functions,
\begin{align}\label{fidph}
\langle{\mathfrak{\bm{f}_1}}|{\mathfrak{\bm{f}_2}}\rangle&=
|\langle{\mathfrak{\bm{f}_1}}|{\mathfrak{\bm{f}_2}}\rangle|
e^{i\varphi},\\
\mathcal{F}_{\rm c}&=\exp\left[-2\langle N\rangle\left(1-\cos\varphi\sqrt{\mathcal{F}_{\rm m}}\right)\right].
\end{align}

\begin{figure}[t]
\begin{center}
\includegraphics[width=0.45\textwidth,height=0.18\textheight]
{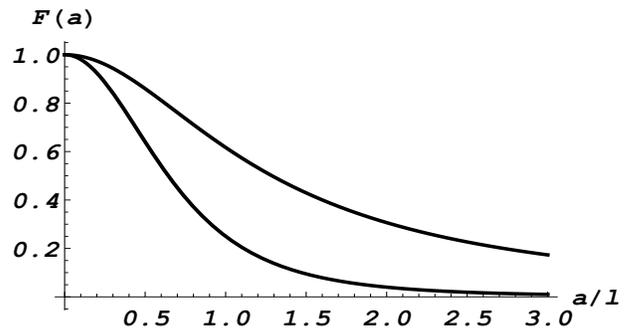}
\caption{Fidelity for momentum wave functions (upper curve) and for position wave functions (lower curve) corresponding to the same photon state .}\label{fig1}
\end{center}
\end{figure}

To compare various measures of fidelity, we have chosen as an example two normalized photon wave functions (with the same helicity) in a simple exponential form:
\begin{align}\label{a}
f_1(\bm k)=\frac{e^{-kl/2}}{\sqrt{4\pi k/l}},\quad f_2(\bm k)=\frac{e^{i k_z a}e^{-kl/2}}{\sqrt{4\pi k/l}}.
\end{align}
Note, that $f_2(\bm k)$ is obtained from  $f_1(\bm k)$ by the translation in the $z$ direction by the distance $a$. In Fig.~\ref{fig1} we plotted fidelity as a function of the distance $a$. Our choice of the function $f_2(\bm k)$ as a function shifted in space  enables us to obtain the information about the photon localization. The faster the fidelity tends to zero with the increase of $a$, the stronger is the localization. In the plot we see that the localization is more stringent in position than in momentum. This was to be expected because the momentum fidelity translated into the position space (\ref{sprf}) has a nonlocal character.
\begin{figure}
\begin{center}
\includegraphics[width=0.45\textwidth,height=0.18\textheight]
{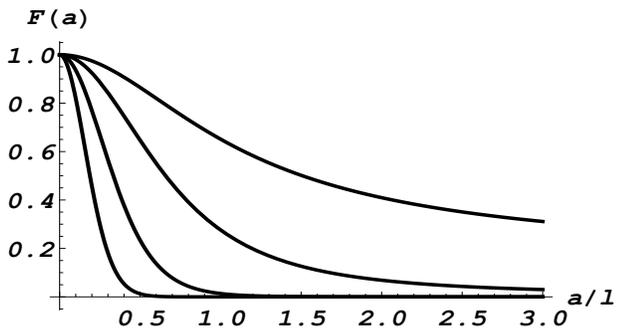}
\caption{Fidelity for the coherent states with the average photon numbers $\langle N\rangle=1$ (uppermost curve), and $\langle N\rangle=3,10,30$.}\label{fig2}
\end{center}
\end{figure}
\begin{figure}[b]
\begin{center}
\includegraphics[width=0.45\textwidth,height=0.18\textheight]
{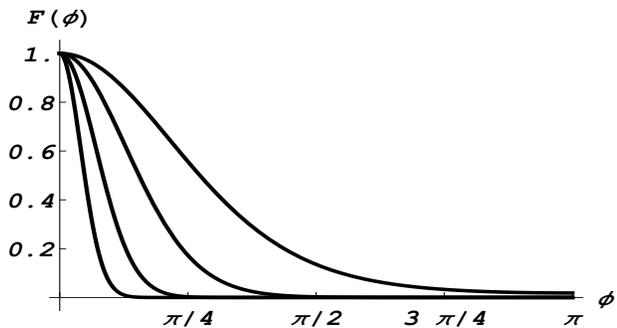}
\caption{Fidelity for two coherent states whose photon wave functions differ only by an overall phase factor $e^{i\varphi}$. Four curves (from top) correspond to: $\langle N\rangle=\{1,3,10,30\}$.}\label{fig3}
\end{center}
\end{figure}

Localizability becomes even more significant for coherent states due to its connection with classicality. Let us, rather arbitrarily, define the extension $s$ of the wave in position space as this value of $a$ for which the fidelity drops to 0.15. This choice leads to the following values in the four cases depicted in Fig.~\ref{fig2}: $s=\{30,1.4,0.6,0.3\}l.$ The larger is $\langle N\rangle$, that is the more classical is the state, the sharper is its localization. For the coherent state with the average number of photons equal to 1, localization is very weak. The curves representing fidelity of this coherent state and the one photon state are similar, as seen by comparing the top curves in Figs.~\ref{fig1} and \ref{fig2}. However, for very large values of $\langle N\rangle$, when the coherent state is almost classical, localization is very strong. Fidelity drops practically to zero for the state shifted in space by a small fraction of its quantum extension $l$.

Finally, we would like to comment on the phase dependence in (\ref{fidph}) in the fidelity between two coherent states. It is a basic tenet of any quantum theory that two wave functions that differ by an overall phase factor describe the same state. However, the fidelity of two coherent states $e^{-\langle N\rangle/2}\exp\left(\langle N\rangle^{1/2}a^\dagger_f\right)|0\rangle$ and $e^{-\langle N\rangle/2}\exp\left(\langle N\rangle^{1/2}e^{i\varphi}a^\dagger_f\right)|0\rangle$ built from {\em the same photons} depends on the phase of the photon wave function, as shown in Fig.~\ref{fig3}.

\section{Conclusions}

The main result of this work is that for photons, unlike for massive particles, fidelities in momentum space and in position space are markedly different. Both descriptions are equally important. The momentum space wave function is perhaps more fundamental due to its direct connection to the Wigner theory of representations \cite{wig}. In turn, the position wave function according to Glauber's theory \cite{roy0} is directly related to photon detection. We have shown how to use fidelity as a measure of photon localizability. We have also calculated fidelity for coherent states of the quantized electromagnetic field and we found very strong dependence on the mean photon number.

\section{Acknowledgment}
We would like to thank the referees for very helpful suggestions that led to significant improvements.

\appendix
\section{Transformation properties of the photon wave functions}

In this Appendix we show how to reconcile the transformation properties of the photon wave functions in momentum representation and in position representation. Under time translation $t\to t-t_0$ and space translation $\bm r\to \bm r-\bm r_0$ the functions $f_\pm({\bm k})$ acquire the same phase factors for both helicities,
\begin{align}\label{trans}
f_\pm(\bm k)\to e^{i\omega t_0}f_\pm(\bm k),\quad f_\pm(\bm k)\to e^{i{\bm k}\cdot{\bm r}_0}f_\pm(\bm k).
\end{align}
These transformations follow from the transformation properties of the electromagnetic field under translations.

In the case of rotations and Lorentz transformations the polarization vector ${\bm e}(\bm k)$ plays an essential role. This vector converts the nonlocal scalar wave function (\ref{wf}) into the electromagnetic field (\ref{genpwf}) with local transformation properties. It follows from the transformation properties of the electromagnetic field that the RS vector transforms under rotations and boosts as follows \cite{pwf}:
\begin{align}\label{trans1}
\Psi'_\pm({\bm r}',t')=\mathcal{O}_\pm\Psi_\pm({\bm r},t),
\end{align}
where $\mathcal{O}_\pm$ are complex orthogonal matrices. With the use of the Fourier representation (\ref{pwf1}), we obtain:
\begin{align}\label{trans2}
\Psi'_\pm({\bm r}',t')=\sqrt{\hbar c}\int\frac{d^3k}{(2\pi)^{3/2}k}k\,{\bm e}_\pm(\bm k)f'_\pm(\bm k)e^{-i\omega t'+i{\bm k}\cdot{\bm r}'}.
\end{align}
Using the transformation properties (\ref{phf}) we obtain:
\begin{align}\label{trans3}
\Psi'_\pm({\bm r}',t')&=\sqrt{\hbar c}\int\frac{d^3k}{(2\pi)^{3/2}k}\nonumber\\
&\times k'{\bm e}_\pm({\bm k}')e^{\pm i\Theta(\bm k)}f_\pm({\bm k})e^{-i\omega t+i{\bm k}\cdot{\bm r}}.
\end{align}
The agreement with the transformation properties (\ref{trans1}) requires that: $k'{\bm e}_\pm({\bm k}')=\mathcal{O}_\pm\,k{\bm e}_\pm({\bm k})e^{\mp i\Theta(\bm k)}$ and we obtain,
\begin{align}\label{form}
e^{-i\Theta(\bm k)}=(k'/k){\bm e}^*({\bm k})\mathcal{O}^T_\pm{\bm e}({\bm k}').
\end{align}
The explicit formula for $\Theta$ in the general case is quite complicated but for some transformations it is relatively simple. For example, for the rotation by angle $\alpha$ around the $y$ axis, we obtain (the wave vector is expressed in spherical coordinates):
\begin{align}\label{roty}
\Theta(\phi,\theta)=\arctan\left(\frac{\sin(\phi)}
{\cot(\alpha)\sin(\theta)-\cos(\phi)\cos(\theta)}\right),
\end{align}
while for the Lorentz boost in the $y$ direction with velocity $v$ the formula is:
\begin{align}\label{boosty}
\Theta(\phi,\theta)=\arctan\left(\frac{v\cos(\phi)\cos(\theta)}
{c\sin(\theta)+v\sin(\phi)}\right).
\end{align}

\end{document}